\documentclass[twocolumn,aps,prl,epsf]{revtex4}
\usepackage{epsfig}
\usepackage{dcolumn}
\usepackage[usenames]{color}
\definecolor{darkorange}{cmyk}{0,1,1,.25}
\def\C{\textcolor{black}}

\newcommand{\be}{\begin{equation}}
\newcommand{\ee}{\end{equation}}
\newcommand{\bea}{\begin{eqnarray}}
\newcommand{\eea}{\end{eqnarray}}
\newcommand{\mbf}[1]{\mbox{\bf{#1}}}

\begin{document}

\title{Gluon Excitations and Quark Chiral Symmetry in the Meson Spectrum:
\\ an Einbein Solution to the Large Degeneracy Problem of Light Mesons}
\author{P. Bicudo}
\affiliation{Dep. F\'{\i}sica and CFTP, Instituto Superior T\'ecnico,
Av. Rovisco Pais, 1049-001 Lisboa, Portugal}
\begin{abstract}
A large approximate degeneracy appears in the light meson spectrum measured at LEAR, suggesting a novel principal quantum number $n+j$ in QCD spectra. 
We recently showed that the large degeneracy could not be understood with state of the art confining and chiral invariant quark models, derived in a truncated Coulomb gauge. To search for a solution to this problem, here we add the gluon or string degrees of freedom. Although independently the quarks or the gluons would lead to a  $2 n +j$ or $2 n +l$ spectrum, adding them may lead to the 
desired $n+j$ pattern. 
\end{abstract}

\maketitle


%
%
%
%
%
%
%
%
%
%

\C{ To understand the large degeneracy of the light meson spectrum
\cite{Bugg:2004xu,Afonin:2006wt,Afonin:2007jd,Glozman:2007ek,Bicudo:2007wt}
observed by the Crystal 
Barrel collaboration at LEAR in CERN, we 
include at the same token both bosonic and fermionic excitations. 
This is relevant for the observation of excited mesons
in the new generation of experiments, say at PANDA at FAIR, and in Lattice QCD.
We first describe the degeneracy, where the meson mass squared $M^2$ 
in the excited meson spectrum is led by the principal quantum number $n+j$. 
Then we review the status of gluon string excitations with a  $M^2$ possibly of the order of 
$ 2 \pi \sigma (2n+l)$. 
Recently we have shown 
\cite{Bicudo:2007wt}
that the quark degrees of freedom are not sufficient to understand the excited light meson spectrum,
finding that the chiral invariant and confining quark model has a $M^2$  of the order of 
$ 2 \pi \sigma (2n+j)$, with a pre-factor is correct for the angular $l$ or $j$ quantum numbers, 
but too large for the radial quantum number $n$. 
Importantly, both gluon and quark excitations have the same scale, with the string tension $\sigma$.
Here we show that arriving at the desired $n+j$ pattern is possible 
if we include the gluon or string degrees of freedom in relativistic quark models with chiral symmetry breaking. 
For simplicity, in this first comprehensive study of gluonic and chiral invariant high spectrum, we use einbeins to add the bosonic excitations to the fermionic excitations. 
}

In the report of Bugg 
\cite{Bugg:2004xu,Afonin:2006wt,Afonin:2007jd,Glozman:2007ek,Bicudo:2007wt}
a large degeneracy emerges from the spectrum of the angularly and radially excited resonances 
produced in $p \bar p$ annihilation by the Crystal Ball collaboration at LEAR 
in CERN
\cite{Aker:1992ny}.
This degeneracy is a larger degeneracy than the chiral degeneracy 
systematically searched by 
Glozman {\em et al} 
\cite{Glozman:2007ek,Wagenbrunn:2007ie,Cohen:2001gb},
Jaffe {\em et al}
\cite{Jaffe:2006jy},  
Afonin
\cite{Afonin:2007mj},
and retrospectively already present  in the light-light spectrum of
Le Yaouanc {\em et al.} 
\cite{Yaouanc}
and the heavy-light spectrum of PB {\em et al.} 
\cite{Bicudo_thesis,Bicudo_hvlt}.
Also notice that a long time ago, Chew and Frautschi remarked the existence
of linear Regge trajectories 
\cite{Chew}
for angularly excited mesons.
A similar linear aligning of excited resonances was also reported for radial excitations
\cite{Anisovich:2001pn}.
Presently the status of the excited meson spectrum approximately generalizes
the Regge trajectories for the total angular momentum $j$,
\be
j+n \simeq \alpha_0 + \alpha M^2
\ee
where $\alpha_0$ is the intercept of the $M^2=0$ axis, 
$\alpha$ is the slope and $n$ is a radial
excitation. The slope for angular
and for radial quantum excitations
\cite{Bugg:2004xu}, 
respectively of 
$0.877 \ \mbox{GeV}^{-2}$ and $ 0.855 \ \mbox{GeV}^{-2} $
are almost identical high in the spectrum 
\C{ and similar to 
$(2 \pi \sigma)^{-1}= 0.84 \ \mbox{GeV}^{-2}$
where the string tension is  $\sigma = 0.19 \ \mbox{GeV}^{2} $.
} 
This is quite remarkable, 
since in nature there are few examples of principal quantum numbers, except
for the coulomb non-relativistic potential, say in the hydrogen atom, or the 
harmonic oscillator non-relativistic potential, say in the vibrational spectrum of 
molecules.

\begin{figure}[t!]
\vspace{-3.7cm}
\hspace{-1cm}
\includegraphics[width=1.2\columnwidth]{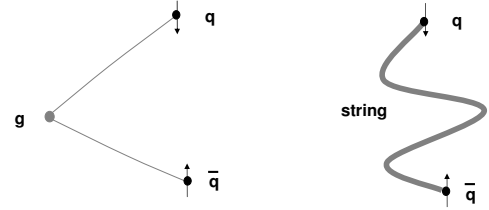}
\vspace{-1.3cm}
\caption{
Gluon (left) and string (right) excitations in a meson. 
}\label{GluonorString}
\end{figure}

We now discuss the gluon or string degrees of freedom in order to include them in the chiral invariant and confining Quark Model. This
is still an open problem, but nevertheless in the end of the letter we will see that it may lead to the observed degeneracy. 
The hybrid excitations of mesons can be addressed with either the
bosonic Nambu-Gotto-Polyakov string fluctuations or with a constituent gluon linked to the quarks by fundamental rigid strings
\cite{Bicudo:2007xp},
as depicted in Fig. \ref{GluonorString}. 
Baker and Steinke
\cite{Baker:2002km}, 
only consider two transverse  modes of the string with 
\C{
rotating
} 
quarks fixed to the ends of the string, resulting in the mass spectrum,
\be
M^2=2 \pi \sigma \left[ n+ l - {1 \over 12}+o\left( 1 \over l \right)\right] \ , 
\label{transverse string}
\ee
led by $n+l$ in units of $2 \pi \sigma$, similar to the spectrum of an open string. This result motivated Afonin 
\cite{Afonin:2007jd}
to propose it as the solution to the large $n+j$ degeneracy, however this does not yet explain the role of the $j$ quantum number, and the chiral partners in the spectrum
\cite{Glozman:2007at}.
Another approach consists in considering a string with static ends, applicable 
to heavy quarks. In this case the static quark-antiquark potential is extended from the linear potential to,
\be
V_{q \bar q}^2(r)= \sigma^2 \, r^2 + 2 \pi \sigma \left( 2n_g +l_g +{\cal E}\right) 
\label{longitudinal too}
\ee
where the pre-factor $2$ of the radial excitation quantum number $n$ is now the double of the
pre-factor $1$  in eq. (\ref{transverse string}) obtained with transverse modes only. $\cal E $ is a constant.
This potential also fits correctly the lattice QCD excitations of the
static quark-antiquark potential measured by Juge, Kuti and Morningstar
\cite{Juge:1997nc}.

We now briefly review how the potential of eq. (\ref{longitudinal too}) has been obtained 
by Buisseret, Mathieu and Semay, utilizing a simple 
\cite{Buisseret:2006sz,Buisseret:2006wc}
constituent quark, antiquark and gluon hamiltonian,
\bea
H & = & 
\sum_{i= q, \bar q , g}
\sqrt{\mbf p_i^2 +m_i^2 }
+ \sum_{j=q, \bar q} 
\sigma  \left| \mbf r_i -  \mbf r_g \right| 
\label{hamiltonian hybrid }
\eea
and applying einbeins, or auxiliary fields,
\bea
\sqrt{\mbf p_i^2 +m_i^2 }
& \simeq &  
\min_{\mu_i}
\left[
{\mbf p_i^2 +m_i^2  \over 2 \mu_i }+ {\mu_i \over 2}
\right]
\nonumber \\
\sigma  \left| \mbf r_i -  \mbf r_g \right| 
& \simeq &  
\min_{\nu_j}
\left[
{ \sigma ^2  \left(\mbf r_i -  \mbf r_g \right) ^2
\over 2 \nu_j}+ {\nu_j \over 2}
\right] \ .
\label{einbein}
\eea
Einbeins were applied to strings by Morgunov, Nefediev and Simonov 
\cite{Morgunov:1999fj}
to include the angular
momentum of rigid strings in the quark confining potentials. Then Buisseret,
Mathieu and Semay utilized the einbeins to also include the string excitations.
The einbeins transform a more difficult
problem into a harmonic oscillator-like Sch\"odinger equation. 
Two independent oscillating modes can be separated, 
one in the usual $\mbf r=\mbf r_q -  \mbf r_{\bar q}$
relative coordinate and another in the $\mbf y={\mbf r_q +  \mbf r_{\bar q}\over 2}- \mbf r_g$ orthogonal coordinate. 
This results in a spectrum, before the einbein minimization, of,
\bea
E&=& {\sigma \over \sqrt{\mu \nu}}(2 n_{q \bar q}+l_{q \bar q}+ 3/2)
+ \sigma \sqrt{ \mu_g+2 \mu \over \mu \nu \mu_g } \times 
\nonumber \\
&&
(2n_g+l_g +3/2) + {m^2 \over \mu}+ \mu +  {m_g^2 \over 2 \mu_g}+{\mu_g \over 2}+ \nu \ .
\label{einbein spectrum}
\eea
In the heavy quark case,  similar to the static limit,  
$\mu \simeq m >> \mu_g$ 
and the minimization of eq. (\ref{einbein spectrum}) leads 
to the string excitations with static quarks as in eq. 
(\ref{longitudinal too}). 
\C{
An important technical detail to arrive at eq. (\ref{longitudinal too})
is the prescription of Buisseret, Mathieu and Semay 
$8 \sigma \to 2 \pi \sigma$ to correct the
einbein enhancement of the radial and angular excitations 
\cite{Buisseret:2006sz,Buisseret:2006wc},
of excited  mesons with a linear potential. Solving the relativistic Schr\"odinger 
equation, the spectrum is led  by $M \to \sqrt{4 \pi \sigma n + 8 \sigma l}$
\cite{Bicudo:2007wt}.
Morgunov, Nefediev  and Simonov
\cite{Morgunov:1999fj}
showed that adding to the angular momentum of
relativistic quarks the angular momentum of a rigid and relativistic rotating string, 
the angular momentum contribution is corrected to  $M \to \sqrt{2\pi \sigma (2 n +  l)}$.
On the other hand the einbein minimization leads to $M \to \sqrt{8 \sigma (2 n +  l)}$.
Therefore using the einbein, minimization together with the 
$8 \sigma \to 2 \pi \sigma$ prescription leads to the spectrum 
$M \to \sqrt{2 \pi \sigma (2 n +  l)}$  of the
solution of the relativistic Schr\"odinger equation for a quark, an antiquark, and a non-vibrating string.
The problem of the pre-factor $2$ of the radial excitation remains, nevertheless that 
is not a result of the harmonic oscillator 
appearing with the einbeins, but a result of the linear potential. The difference 
between the spectra produced with the harmonic oscillator and the linear potentials
is the exponent of the principal quantum number in the spectrum. In particular a linear potential
contributes to the spectrum as $M^2 \to 2 \pi \sigma (2 n + l)$
to linear Regge spectra. This differs from the Harmonic Oscillator spectra 
$M \to \omega (2 n + l)$.
Also notice that PB 
\cite{Bicudo:2007wt}
tested enlarging the anzatse to the class of different power law confining
potentials but this did not solve the problem of the principal quantum number,
and it also may produce non-linear Regge trajectories. So one would better maintain the
potential linear, and look for another solution to the pre-factor $2$ problem. 
}

Importantly, eqs.  (\ref{transverse string}), (\ref{longitudinal too}) 
and and the Regge behaviour 
$M \to \sqrt{2 \pi \sigma (2 n +  l)}$ of relativistic (but spinless) quarks 
suggest that a similar quantitatively correct angular 
Regge slope $\alpha = (2 \pi \sigma)^{-1}$
is possible with both quark and gluonic excitations.
However these eqs. still lack the quark spin degrees of freedom,
in particular they miss the insensitivity to chiral symmetry breaking
observed in the large degeneracy of the excited meson spectrum.
To solve this $j$ quantum number problem, it is unavoidable to
work with confined and relativistic quarks with spontaneous 
chiral symmetry breaking. 
\C{
Notice that the spontaneous 
chiral symmetry breaking does not affect the
Regge slopes, but it does correct the intercept, in particular 
the $\pi$ meson is massless in the chiral limit, and the 
$\rho$ meson mass is reproduced. The $\rho$ sits in the leading
Regge trajectory and determines its intercept.
High in the
spectrum the good angular quantum number is not the orbital one
$l$ but the total one $j$.
}
The necessary techniques were
developed by Le Yaouanc, Oliver, Ono, Pene and Raynal
\cite{Yaouanc},
and by PB and Ribeiro
\cite{Bicudo_thesis}, 
and extended to hybrids by Llanes-Estrada, General and Cotanch
\cite{LlanesEstrada:2000hj,General:2006ed}. 
But in this letter we neglect the spin contribution of the gluon to
the hamiltonian, since the fine structure of the static potential,
also measured in Lattice QCD
\cite{Juge:2002br},
suggests that the spin $s_g$ degree of freedom of the gluon 
only contributes at small distances, 
and $s_g$ is not a leading term for excited mesons. 
The hamiltonian  for the quarks is, 
\begin{eqnarray}
&&H=\int\, d^3x \left[ \psi^{\dag}( x) \;(m_0\beta -i{\vec{\alpha}
\cdot \vec{\nabla}} )\;\psi( x)\;+
{ 1\over 2} g^2 \int d^4y\, \
\right.
\nonumber \\
&&
\overline{\psi}( x)
\gamma^\mu{\lambda^a \over 2}\psi ( x)  
V_{\mu \, \nu}^{a \, b}(\mbf x-\mbf y)\, \overline{\psi}( y)
\gamma^\nu{\lambda^b \over 2}
 \psi( y) \  ,
\label{hamilt}
\end{eqnarray}
where it is standard in the Coulomb Gauge to use a density-density colour octet potential
$V_{\mu \, \nu}^{a \, b}(\mbf r)= \delta_{\mu 0} \delta _{\nu 0}\delta^{ab} (- 3/16) V(r)$, 
with a funnel potential for the spatial function 
$V( r)= {\alpha \over r} + v_0 + \sigma r $. 
\C{
Notice that for excited states we can neglect the Coulomb part $ {\alpha \over r}$ of the potential
\cite{Bicudo:2008kc}
since it would have little effect in the excited spectrum because the quark wavefunctions have
a greater radius mean square. Moreover, when chiral symmetry breaking occurs,
the constant term $v_0$ is cancelled in the spectrum by an opposite term present in the quark
self-energy. Thus for the study of quark degrees of freedom in excited mesons it is sufficient
to consider the linear potential $\sigma r $ only. 
}

\C{
Here we go beyond these previous 
studies of mesons with chiral invariant quarks, including in this framework
the gluon or string excitations,
as in eq. (\ref{hamiltonian hybrid }), 
considering a vanishing current quark mass $m_q=0$.
We also apply the einbein technique to simplify the resulting three-body problem
as in eq. (\ref{einbein}). This allows us to separate the two coordinates $\mbf r$ and $\mbf y$.
In what concerns the relative quark-antiquark $\mbf r$ coordinate, we address chiral symmetry 
breaking the bethe Salpeter equation for the quark
and the antiquark with the framework already developed for mesons.
We now review the technical steps of this framework, before including the
gluon or string degree of freedom, corresponding to the coordinate $\mbf y$.
}
The boundstate equations are derived translating 
the relativistic invariant Dirac-Feynman propagators
\cite{Yaouanc}, 
in the quark and antiquark Bethe-Goldstone 
propagators
\cite{Bicudo:2007wt,Bicudo_scapuz},
used in the formalism of non-relativistic quark models,
\FL
\begin{eqnarray}
{\cal S}_{Dirac}(k_0,\vec{k})
&&= {i \over k_0 -\sqrt{k^2+m_c^2(k)} +i \epsilon} \
\sum_su_su^{\dagger}_s \beta
\nonumber \\
&& \ \  - {i \over -k_0 -\sqrt{k^2+m_c^2(k)} +i \epsilon} \
\sum_sv_sv^{\dagger}_s \beta \ ,
\nonumber \\
u_s({\bf k})=&& \left[
\sqrt{ 1+S \over 2} + \sqrt{1-S \over 2} \widehat k \cdot \vec \sigma \gamma_5
\right]u_s(0)  \ ,
\nonumber \\
v_s({\bf k})=&& \left[
\sqrt{ 1+S \over 2} - \sqrt{1-S \over 2} \widehat k \cdot \vec \sigma \gamma_5
\right]v_s(0)  \ ,
\label{propagators}
\end{eqnarray}
where it is convenient to define different functions of the constituent quark mass $m_c(k)$, $\varphi (k)= \arctan{m_c (k) \over k}$, $ S(k)= \sin \varphi (k)$,   $ C= \cos \varphi (k)$ and ${\cal G}(k)= 1 - S(k) $.
In the non condensed vacuum, $m_c(k)$ equals the bare quark mass $m_0$ .
In the physical vacuum, the constituent quark mass $m_c(k)$ is a variational function which is determined by the mass gap equation. 
But here we do not detail it since the constituent mass is only crucial
for the groundstate mesons, but not for the higher states in the spectrum
where it can be neglected when compared to the momentum of the quark. 
%
%
\begin{table}[t]
\begin{ruledtabular}
\begin{tabular}{c|c}
 & $V^{++}_{HO}=V^{--}_{HO}$  \\ \hline
spin-indep. & $- {d^2 \over dk^2 } + { {\bf L}^2 \over k^2 } + 
{1 \over 4} \left( {\varphi'_q}^2 + {\varphi'_{\bar q}}^2 \right) 
+ {  1 \over k^2} \left( {\cal G}_q +{\cal G}_{\bar q}  \right) -U $  \\ 
spin-spin & $ {4 \over 3 k^2} {\cal G}_q {\cal G}_{\bar q} {\bf S}_q \cdot {\bf S}_{\bar q} $  \\ 
spin-orbit & $ {1 \over  k^2} \left[ \left( {\cal G}_q + 
{\cal G}_{\bar q} \right) \left( {\bf S}_q +{\bf S}_{\bar q}\right) 
+\left( {\cal G}_q - {\cal G}_{\bar q} \right) \left( {\bf S}_q -{\bf S}_{\bar q}\right)  \right]
\cdot {\bf L} $  \\ 
tensor & $ -{2 \over  k^2} {\cal G}_q {\cal G}_{\bar q} 
\left[ ({\bf S}_q \cdot \hat k ) ({\bf S}_{\bar q} \cdot \hat k )
-{1 \over 3} {\bf S}_q \cdot {\bf S}_{\bar q} \right] $ \\ \hline
 & $V^{+-}_{HO}=V^{-+}_{HO}$ \\ \hline
spin-indep. & $0$  \\ 
spin-spin & $ -{4 \over 3} \left[ {1\over 2} {\varphi'_q} {\varphi'_{\bar q}} + 
{1\over k^2} { C}_q { C}_{\bar q}  \right]
{\bf S}_q \cdot {\bf S}_{\bar q} $  \\ 
spin-orbit & $0$  \\ 
tensor & $ \left[ -2 {\varphi'_q} {\varphi'_{\bar q}} + 
{2\over k^2}{  C}_q { C}_{\bar q}  \right]
\left[ ({\bf S}_q \cdot \hat k ) ({\bf S}_{\bar q} \cdot \hat k )
-{1 \over 3} {\bf S}_q \cdot {\bf S}_{\bar q} \right] $
\end{tabular}
\end{ruledtabular}
\caption{\label{spin dependent} 
The positive and negative energy spin-independent, spin-spin, spin-orbit and tensor 
potentials, computed exactly in the framework of the simple density-density harmonic oscillator. For a detailed derivation,  see reference 
\cite{Bicudo:2007wt,Bicudo_scapuz}.
}
\end{table}
The Salpeter-RPA equations for a meson (a colour singlet
quark-antiquark bound state) can be derived from the Lippman-Schwinger
equations for a quark and an antiquark, or replacing the propagator
of eq. (\ref{propagators}) in the Bethe-Salpeter equation. 
For a detailed derivation of the boundstate equations see reference 
\cite{Bicudo_scapuz}. One gets,
\FL
\begin{eqnarray}
\label{homo sal}
R^+(k,P) &=& { u^\dagger(k_1) \chi(k,P)  v(k_2) 
\over +M(P)-E(k_1)-E(k_2) }
\nonumber \\
{R^-}^t(k,P) &=& { v^\dagger(k_1) \chi(k,P) u(k_2)
\over -M(P)-E(k_1)-E(k_2)}
\nonumber \\
\chi(k,P) &=&
\int {d^3k' \over (2\pi)^3} V(k-k') \left[ 
u(k'_1) R^+(k',P)v^\dagger(k'_2) \right.
\nonumber \\
&&\left. +v(k'_1){ R^-}^t(k',P) u^\dagger(k'_2)\right] 
\label{Bethe-Salpeter}
\end{eqnarray}
where $V(k)$ is the Fourier transform of the
potential in eq. (\ref{hamilt}), and where 
$k_1=k+{P \over 2} \ , \ k_2=k-{P \over 2}$ and $P$ is
the total momentum of the meson.
Eq. (\ref{Bethe-Salpeter}) includes both the
Salpeter-RPA equations of PB et al. 
\cite{Bicudo_thesis}
and of Llanes-Estrada et al. 
\cite{Llanes-Estrada_thesis}
and the Salpeter equations of Le Yaouanc et al.
\cite{Yaouanc}. 
Substituting $\chi$ we get the equation for the positive energy
$R^+$ and negative energy $R^-$ radial wavefunctions.
This results in four potentials 
$V^{\alpha \beta}$ with $\alpha=\pm$, 
respectively coupling $\rho^\alpha=k \, R^\alpha$ to 
$\rho^\beta$, in the boundstate Salpeter equation,
\begin{equation}
\left\{
\begin{array}{rrr}
\left(2 T + V^{++} \right) \, \rho^+ \  + & V^{+-} \ \ \rho^- =& M \rho^+ \\
V^{-+} \ \  \rho^+ \ + & \left( 2 T + V^{--} \right) \, \rho^- =& - M \rho^- 
\end{array}
\right. .
\label{decoupled}
\end{equation}

We now apply some educated approximations to arrive at an analytical 
expression for the excited meson spectrum. 
As in eq. (\ref{einbein}), we use einbeins
and this avoids the problem of computing the matrix elements of the
linear potential, actually we only need to know the matrix elements 
of the harmonic oscillator potential 
$V^{\alpha \beta}_{HO}$ as shown 
\cite{Bicudo:2007wt}
in Table \ref{spin dependent},
and to minimize each energy in the spectrum with regards to $\mu$, $\nu$
and $\mu_g$.
Importantly, the potentials $V^{++}=V^{--}$ and $V^{+-}=V^{-+}$ include 
the usual spin-tensor potentials
\cite{Bicudo:2007wt,Bicudo_scapuz}, 
produced by the Pauli $\vec \sigma$ matrices in the spinors of eq. (\ref{propagators}).
They are detailed explicitly in Table \ref{spin dependent}.
Moreover we are interested in highly excited states, 
where both $\langle r \rangle $ and $\langle k \rangle $ are large
due to the virial theorem, 
and thus we consider the limit where ${m_c \over k} \rightarrow 0$ as in PB, Cardoso, Van Cauteren and Llanes
\cite{Bicudo:2009cr}.    
This implies that in the potentials listed in Table \ref{spin dependent},
we may simplify $\varphi'(k)\rightarrow 0$, 
${\cal C}(k)\rightarrow 1$ and 
${\cal G}(k)= 1 - S(k)  \rightarrow 1 $. 
We also notice that the relativistic equal time equations have twice as 
many coupled equations as does the Schr\"odinger equation. But the negative energy component 
$\rho^-$ is smaller than the positive energy component $\rho^+$ by a factor of the
order of $\sqrt{\sigma}/M$. Thus when the meson mass $M$ is large, 
and this is the case for the excited mesons, the negative energy components can be neglected and the Salpeter equation simplifies to a 
Schr\"odinger equation, where
the relevant potential $V^{++}$ becomes,
\bea
V^{++}_{HO}&=&- {d^2 \over dk^2 } + { {\mbf L}^2 \over k^2 } 
+ {  2 \over k^2} 
+{2 \over  k^2} \left( {\mbf S}_q +{\mbf S}_{\bar q}\right) 
\cdot {\mbf L} 
\nonumber \\
&& +{4 \over 3 k^2}  {\mbf S}_q \cdot {\mbf S}_{\bar q} 
 -{2 \over  k^2} 
\left({\mbf S}_q \cdot \hat k  \  {\mbf S}_{\bar q} \cdot \hat k 
-{1 \over 3} {\mbf S}_q \cdot {\mbf S}_{\bar q} \right)
\nonumber \\
& = &
- {d^2 \over dk^2 } + { {\mbf J}^2 \over k^2 } + {{\cal E}' \over k^2}
\label{spinsimple}
\eea
the dependence on $\mbf L^2$ is traded by a dependence
on $\mbf J^2=\mbf L^2+ 2 \mbf L \cdot \mbf S + \mbf S^2 $. The other 
remaining potentials are the spin-spin and tensor
potentials, both much smaller than $k^2$ or than $J^2$,
and they simply contribute to a Coulomb-like potential ${{\cal E}' \over k^2}$.

Including the gluonic coordinate $\mbf y$ and Fourier transforming it as well,
removing the Coulomb terms un-affecting the excited spectrum, we get
simple Harmonic Oscillator terms. Substituting
the Harmonic Oscillator matrix elements, we arrive at the meson spectrum,
\bea
M&=& {\sigma \over \sqrt{\mu \nu}}(2 n_{q \bar q}+j_{q \bar q}+ 3/2)
+ \sigma \sqrt{ \mu_g+2 \mu \over \mu \nu \mu_g } \times 
\nonumber \\
&&
(2n_g+l_g +3/2) + \mu +  {m_g^2 \over 2 \mu_g}+{\mu_g \over 2}+ \nu \ ,
\label{new einbein spectrum}
\eea
before minimization,
similar to eq. (\ref{einbein spectrum}) except that the quark orbital angular momentum $l_{q \bar q}$ is replaced
by the total quark angular momentum $j_{q \bar q}$.
Eq (\ref{new einbein spectrum}) is hard to minimize analytically, except in two different limits,
the limit of a large effective gluon mass, and the limit of a small number of gluon excitations.
Then we get,
\be
E\simeq \sqrt{ 2 \pi \sigma 
\left({ \cal N}_{q \bar q} + {\cal N}_g 
\right)} + m_g \ ,
\label{spectrum with spin, large mg}
\ee
where we denote ${\cal N}_{q \bar q}= 2 n_{q \bar q} + j_{q \bar q} + 3/2$
and ${\cal N}_g= 2n_g+l_g +3/2$.
In the case of a vanishing effective gluon mass or of a significant ${\cal N}_g$ we have to resort to
a numerical minimization of eq. ({new einbein spectrum})
(again with the $8 \sigma \to 2 \pi \sigma$ prescription) 
and we get the fit
\bea
M &\simeq &
\sqrt{   2 \pi \sigma \left( 
{ \cal N}_{q \bar q} + {\cal N}_g 
\right) }
 \left[ 1.29
+ 
0.18{ { \cal N}_{q \bar q} - {\cal N}_g  \over { \cal N}_{q \bar q} + {\cal N}_g }
\right.
\nonumber \\
&&
\left.+ 
0.08 
\left( 
{ { \cal N}_{q \bar q} - {\cal N}_g  \over { \cal N}_{q \bar q} + {\cal N}_g  } 
\right)^2 
\right] \ ,
\label{numerical minimization}
\eea
in terms of a dominant principal number ${ \cal N}_{q \bar q} + {\cal N}_g$, 
and a small contribution from a second number ${ \cal N}_{q \bar q} - {\cal N}_g$.
Notice that the principal quantum number is in general much larger than the
second number, due to the zero point energy of the Harmonic Oscillator.
The simplest of all these cases is the one of a massless gluon and of few gluonic
excitations, producing 
\be
M \simeq 
\sqrt{ 
 2 \pi \sigma 
\left( { \cal N}_{q \bar q} + {\cal N}_g \right)
} \ ,
\label{simplified spectrum}
\ee 
also leading to linear Regge trajectories.

\begin{figure}[t!]
\vspace{-2.1cm}
\hspace{-1.80cm}
\includegraphics[width=1.2\columnwidth]{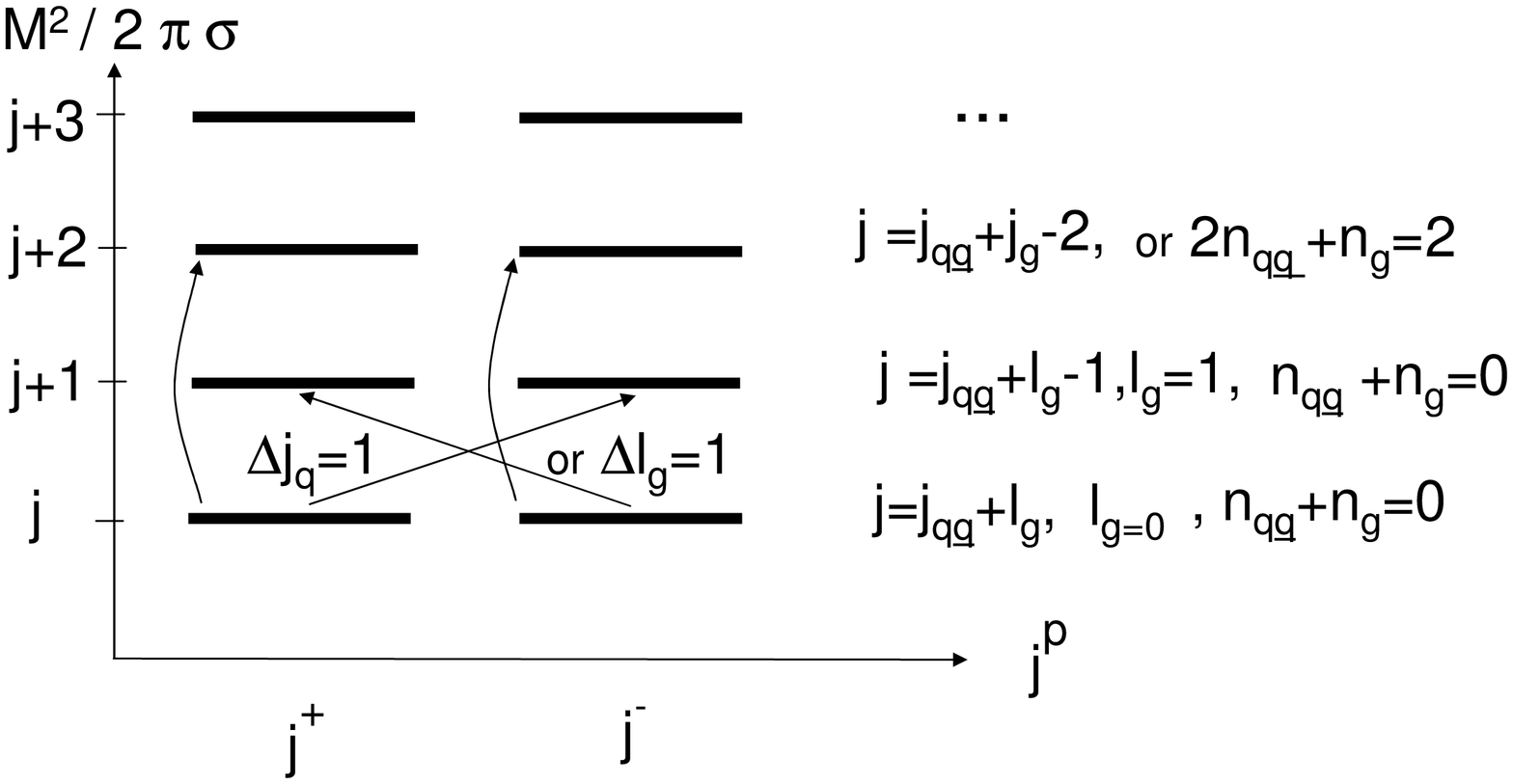}
\vspace{-1.1cm}
\caption{
Mass squared $M^2$, as a function of $j^{P}$, of radial-like excitations in $2 \pi \sigma$ 
units starting from the groundstates in the total angular momentum $j$ shell.
}\label{spectrum}
\end{figure}

Thus in any of the cases considered we find, as a good approximation 
that a principal quantum number ${ \cal N}_{q \bar q} + {\cal N}_g$
dominates the excited light meson 
spectrum.  We now show 
that the principal quantum number 
${ \cal N}_{q \bar q} + {\cal N}_g =2 n_{q \bar q} + j_{q \bar q} + 2 \, n_g +l_g  + 6 $
can be simplified to a form $ n +j$. This
is quite interesting since it allows us to understand the experimental
spectrum with both the quark quantum numbers and the gluon quantum numbers!
Obviously, the trajectories in $\mbf J_{total}=\mbf J_{q \bar q} + \mbf L_g$
have the correct slope of $ (2 \pi \sigma)^{-1}$. Thus we only need to show
that the radial-like trajectories, i. e. the ones that maintain $j_{total}$ and the parity $P$,
also have the same slope, even in the less favourable case where the gluonic radial
excitation pre-factor is $2$ and not $1$ as in eq. (\ref{transverse string}).
As in the radial trajectories of Bugg
\cite{Bugg:2004xu}, 
\C{
let us consider a radial trajectory starting with mesons which already have angular quark 
excitations $ j_{q \bar q} > 0$. 
Due to the chiral symmetry insensitivity we start with a degenerate parity doublet, including a meson
with $j_{total}=j_{q \bar q}+j_g$ and $P=+$  and another 
with $j_{total}=j_{q \bar q}+j_g$ and $P=-$, i. e. respectively with $j ^P= j_{total}^+$ and 
 $j_{total}^-$.
The groundstate, in the leading trajectory, has with no
hybrid excitation, i. e. it has $n_g=0$ and with $l_g=0$. 
Then, increasing $ l_g$ by one unit and summing it to the quarks angular momentum
$j_{q \bar q}$ 
with the triangular relation  $ \left| j_{q \bar q}  -l_g \right|<j_{total}< j_{q \bar q}  +l_g $
it is clearly possible to maintain the initial $j_{total} = j_{q \bar q} $. 
Although the gluon angular momentum does invert the parity of the initial state,
as depicted in Fig. \ref{spectrum}, since we start with a parity doublet, this 
gluonic excitation
\be
\Delta j =0\ , \ \ \Delta M^2= 2 \pi \sigma \ \ ,
\ee
leads to another parity doublet. 
This doublet is a radial-like excitation of the one we started from.
But, importantly, the $M^2$ splitting for this radial-like 
excitation is only $ 2 \pi \sigma$. 
The next excitation, with $ \Delta M^2= 2 \times 2 \pi \sigma$ is easier 
to get since it can be intrinsically radial, 
with  $\Delta (2 n_{q \bar q} +  c \, n_g)=2$. And so forth, with the interplay
of $ \mbf J_{q \bar q} $ and $ \mbf L_g  $ we are able to produce new radial-like
excitations, in the middle of the purely radial excitations.
Thus the Regge slope for the radial-like excitations 
$ \alpha=(2 \pi \sigma)^{-1}$ is the same one of the angular excitations.
}

\C{
To conclude, we address a novel large degeneracy observed in quantum physics,
with a principal quantum number $n+j$ in the excited light meson spectrum.
With the interplay of quark excitations insensitive to chiral symmetry and 
of gluon string excitations we show that the degeneracy may be explained.
We simplify the quark with strings problem to a 3-body problem with
two chiral invariant quarks and a gluon.
An exact numerical solution of this three-body problem 
would require a similar effort to the one necessary to
study other three-body systems of chiral invariant quarks and gluons
\cite{Bicudo:2009cr,LlanesEstrada:2005jf},
e. g. 9 dimensional spatial integrals, the sum of 6 spin indices, 
and the use of a variational  
basis with hundreds of multidimensional orthogonal functions,
implying man-years of coding
and debugging together with the use of a supercomputer for several months.
Thus for this first study,  and to to arrive at simple and clear analytical mass formulas,
we use the einbein technique. 
As a by-product of our research, we find that the einbein technique is applicable
to the quark models with chiral symmetry breaking. For instance, eq. (\ref{simplified spectrum}),
with the quark degrees of freedom only, should provide a good approximation 
to the exact numerical solution of Llanes-Estrada and Cotantch 
\cite{Llanes-Estrada_thesis}
and of Glozman and Wagenbrunn
\cite{Wagenbrunn:2007ie}.
}

\C{ {\bf Note added in proof -} Very recently Bugg re-analysed Cristal Barrel data 
at LEAR 
\cite{Bugg:2009me}
and his spectrum shows clearly the large degeneracy, where the angular
excitations have the same scale as the radial-like excitations.
However he found no evidence for parity doublets in the leading Regge Trajectory,
but only in daughter trajectories.
More experimental, computational and theoretical works on the light meson
spectrum are urgent.
}
 
%

\acknowledgements

PB tanks the colleagues at the 2009 Schladming Winter School for motivating the effort necessary to complete this paper. 
PB was supported by the FCT grants POCI/FP/81913/2007, POCI/FP/81933/2007 and 
CERN/FP/83582/2008.


\end{document}